\documentclass[11pt]{amsart}
\usepackage[leqno]{amsmath}
\usepackage{epsfig}
\usepackage{graphicx}
\usepackage{cite}
\usepackage[utf8]{inputenc}
\usepackage{algorithm, algorithmic}
\usepackage{subfigure}
\usepackage{epsfig}
\usepackage{color}
\usepackage{amsmath}
\usepackage{amssymb}
\usepackage{graphicx,color, algorithmic, algorithm}

\newtheorem{theorem}{Theorem}

\begin{document}

\thispagestyle{empty}

	\centerline{\Large\bf Embedding into the rectilinear plane in optimal $O(n^2)$ time}

	\vspace{6mm}

	\centerline{{\sc Nicolas Catusse, Victor Chepoi,} and {\sc Yann Vax\`es}}

	\vspace{3mm}
	\medskip
	\centerline{Laboratoire d'Informatique Fondamentale,}
	\centerline{Universit\'e d'Aix-Marseille,}
	\centerline{Facult\'e des Sciences de Luminy,} \centerline{F-13288
	Marseille Cedex 9, France} \centerline{$\{$catusse,chepoi,vaxes$\}$@lif.univ-mrs.fr}

	\vspace{15mm}
	\begin{footnotesize} \noindent {\bf Abstract.}
 In this paper, we present
an optimal $O(n^2)$ time algorithm for deciding if a metric space
$(X,d)$ on $n$ points can be isometrically embedded into the plane
endowed with the $l_1$-metric. It improves the
$O(n^2\log^2n)$ time algorithm of J. Edmonds (2008). Together with
some ingredients introduced by Edmonds, our algorithm uses the
concept of tight span and the injectivity of the $l_1$-plane. A
different $O(n^2)$ time algorithm was recently proposed by D.
Eppstein (2009).
\end{footnotesize}

\section{Introduction}

Deciding if a finite metric space $(X,d)$ admits an isometric embedding or an embedding with a small distortion
into a given geometric space (usually ${\mathbb R}^k$ endowed with some norm-metric) is a classical question
in distance geometry  which has some applications in theoretical computer science, visualization, and data analysis.
The first question can be answered in polynomial time
if ${\mathbb R}^k$ is endowed with the Euclidean metric due to classical results of Menger and Sch\"onberg \cite{DeLa}.
On the other hand, by a result of Frechet \cite{DeLa}, any metric space can be isometrically embedded into some
${\mathbb R}^k$ with the $l_{\infty}$-metric. However, it is NP-hard  to decide if a metric space isometrically
embeds into some ${\mathbb R}^k$ endowed with the $l_1$ (alias rectilinear or Manhattan) metric \cite{AvDe,DeLa}. More recently,
Edmonds \cite{Ed} established that it is even NP-hard to decide if a metric space embeds into ${\mathbb R}^3$ with
$l_{\infty}$-metric (a similar question for ${\mathbb R}^3$ with
$l_1$-metric is still open).  In case of ${\mathbb R}^2,$ both $l_1$- and $l_{\infty}$-metrics are equivalent because the
second metric can be obtained from the first one by a rotation of the plane by $45^{\circ}$  and then by a shrink by a factor
$\frac{1}{\sqrt{2}}.$ The embedding problem for the rectilinear plane was investigated in the papers \cite{BaCh_six,MaMa}, which ultimately
show that a metric space $(X,d)$ embeds into the $l_1$-plane if and only if any subspace
with at most six points does \cite{BaCh_six}  (a similar result for embedding into the $l_1$-grid was obtained in \cite{BaCh_grid}).
As a consequence, it is possible to decide in polynomial time if a finite metric space embeds into the $l_1$-plane.
Edmonds \cite{Ed} presented an $O(n^2\log^2n)$ time algorithm for this problem and very recently we learned that
Eppstein \cite{Epp} described an optimal $O(n^2)$ time algorithm (for earlier algorithmic results, see also \cite{ChTr}).
In this note, independently of \cite{Epp}, we describe a simple and optimal algorithm for this embedding problem,
which is different from that of \cite{Epp}.

We conclude this introductory section with a few definitions. In the
sequel, we will denote by $d_1$ or by $||\cdot||_1$ the $l_1$-metric
and by $d_{\infty}$ the $l_{\infty}$-metric. A metric space $(X,d)$
is {\it isometrically embeddable} into a host metric space $(Y,d')$
if there exists a map $\varphi: X\rightarrow Y$ such that
$d'(\varphi(x),\varphi(y))=d(x,y)$ for all $x,y\in X.$ In this case
we say that $X$  is a subspace of $Y.$ A {\it retraction} $\varphi$
of  a metric space $(Y,d)$ is an idempotent nonexpansive mapping of
$Y$ into itself, that is, $\varphi^2=\varphi:Y\rightarrow Y$ with
$d(\varphi (x),\varphi (y))\le d(x,y)$ for all $x,y\in Y.$ The
subspace  of $Y$ induced by the image of $Y$ under $\varphi$ is
referred to as a {\it retract} of $Y.$ Let $(X,d)$ be a metric
space.  The {\it (closed) ball} and the {\it sphere} of center $x$
and radius $r$ are the sets $B(x,r)=\{ p\in X: d(x,p)\le r\}$ and
$S(x,r)=\{ p\in X: d(x,p)=r\},$ respectively. The \emph{interval}
between two points $x,y$ of $X$ is the set $I(x,y)=\{ z\in X:
d(x,y)=d(x,z)+d(z,y)\}.$  Any ball of $({\mathbb R}^k,d_{\infty})$
is an axis-parallel cube. A subset $S$ of $X$ is \emph{gated} if for
every point $x\in X$ there exists a (unique) point $x'\in S,$ the
\emph{gate} of $x$ in $S$,  such that $x'\in I(x,y)$ for all $y\in
S$ \cite{DrSch}. The intersection of gated sets is also gated.  A
{\it geodesic} in a metric space is the isometric image of a line
segment. A metric space is called {\it geodesic} (or {\it
Menger-convex}) if any two points are the endpoints of a geodesic.

For a point $p$ of ${\mathbb R}^2,$ denote by $Q_1(p),\ldots,Q_4(p)$ the
four quadrants of ${\mathbb R}^2$ defined by the vertical and horizontal
lines passing via the point $p$ and labeled counterclockwise.
Any interval $I_1(x,y)$ of the rectilinear plane $({\mathbb R}^2,d_1)$ is an
axis-parallel rectangle which can be
reduced to a horizontal or vertical segment. Any ball of $({\mathbb
R}^2,d_1)$ is a lozenge obtained from an axis-parallel square by a
rotation by $45^{\circ}$ degrees. In the rectilinear plane, any halfplane defined
by a vertical or a horizontal line is gated. As a consequence,
axis-parallel rectangles, quadrants, and strips of  $({\mathbb
R}^2,d_1)$ are gated as intersections of such halfplanes.

\section{Tight spans}

A metric space $(X,d)$ is called {\it hyperconvex} (or {\it
injective})  \cite{ArPa,Is} if any family of closed balls
$B(x_i,r_i)$ with centers $x_i$ and radii $r_i,$ $i\in I,$
satisfying $d(x_i,x_j)\le r_i+r_j$ for all $i,j\in I$ has a nonempty
intersection, that is, $(X,d)$ is a geodesic space such that the
closed balls have the Helly property. Since the closed
balls of $({\mathbb R}^k,d_{\infty})$ are axis-parallel boxes, the
metric spaces $({\mathbb R}^k,d_{\infty})$ and $({\mathbb R}^2,d_1)$
are hyperconvex. It is well known \cite{ArPa} that $(X,d)$ is
hyperconvex iff it is an absolute retract, that is,
$(X,d)$ is a retract of every metric space into which it embeds
isometrically.  As shown by Isbell \cite{Is} and Dress \cite{Dr},
for every metric space $(X,d)$ there exists the smallest injective
space $T(X)$ extending $(X,d),$ referred to as the {\it injective
hull} \cite{Is}, or {\it tight span} \cite{Dr} of $(X,d).$ The tight
span of a finite metric space $(X,d)$ can be defined as follows. Let
$T(X)$ be the set of functions $f$ from $X$ to ${\mathbb R}$ such
that

\medskip
(1) for any $x,y$ in $X,$ $f(x)+f(y)\ge d(x,y),$ and

(2) for each $x$ in $X,$ there exists $y$ in $X$ such that $f(x)+f(y)=d(x,y).$

\medskip
One can interpret $f(x)$ as the distance from $f$ to $x$. Then (1)
is just the triangle inequality.  Taking $x=y$ in (1), we infer that
$f(x)\ge 0$ for all $x\in X.$  The requirement (2) states that
$T(X)$ is minimal, in the sense that no value $f(x)$ can be reduced
without violating the triangle inequality. We can endow $T(X)$ with
the $l_{\infty}$-distance: given two functions $f$ and $g$ in
$T(X),$ define $\rho(f,g)=\max |f(x)-g(x)|.$ The resulting metric
space $(T(X),\rho)$ is injective and $(T(X),\rho)$ is called the
{\it tight span} of $(X,d).$ There is an isometric embedding of $X$
into its tight span $T(X)$. Moreover, {\it any isometric embedding
of $(X,d)$ into an injective metric space $(Y,d')$ can be extended
to an isometric embedding of $(T(X),\rho)$ into $(Y,d'),$} i.e.,
$(T(X),\rho)$ is the smallest injective space into which $(X,d)$ embeds isometrically.

\begin{figure}
\begin{center}
\begin{tabular}{cc}
    \includegraphics[scale=0.5]{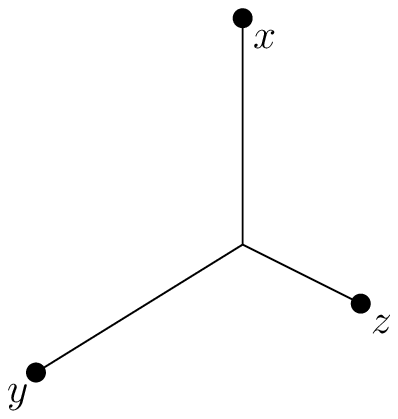} & \hspace{2cm}
\includegraphics[scale=0.5]{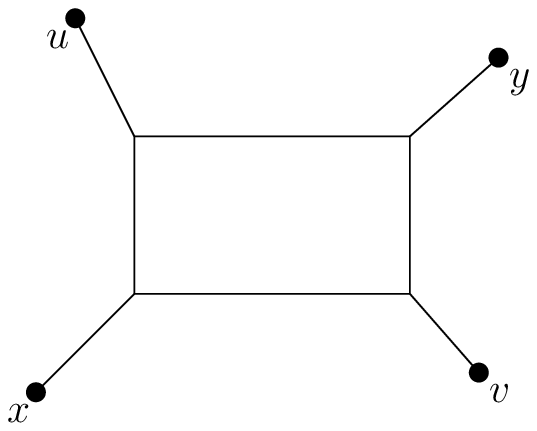} \vspace{-3mm}\\
(a) & \hspace{2cm}(b) \\
\end{tabular}
\end{center}
\caption{Tight span of 3- and 4-point metric space.}
\label{TS_3_4}
\end{figure}

\begin{figure}
\begin{center}
\begin{tabular}{ccc}
    \includegraphics[scale=0.5]{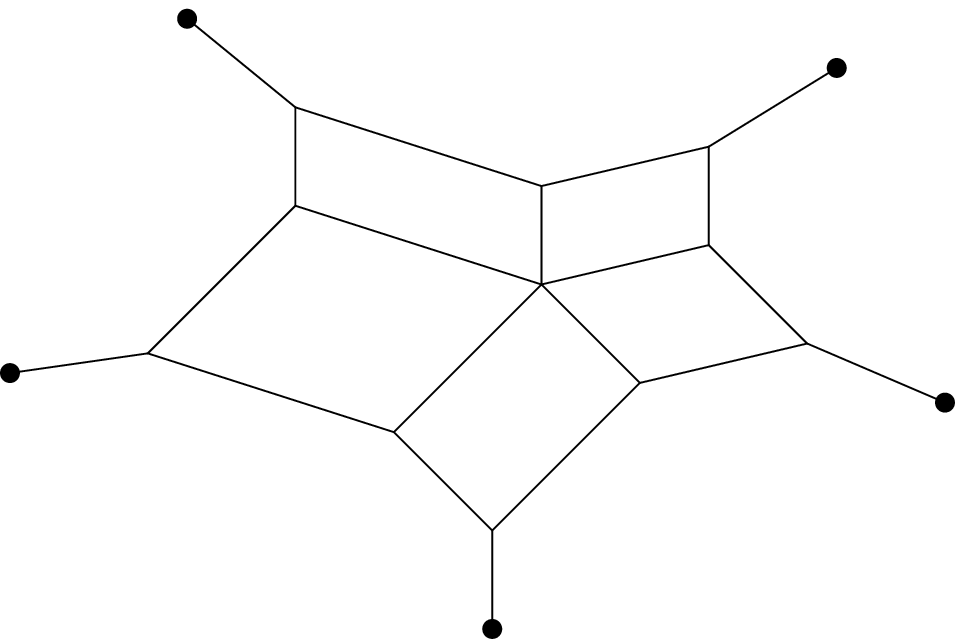} &
    \includegraphics[scale=0.5]{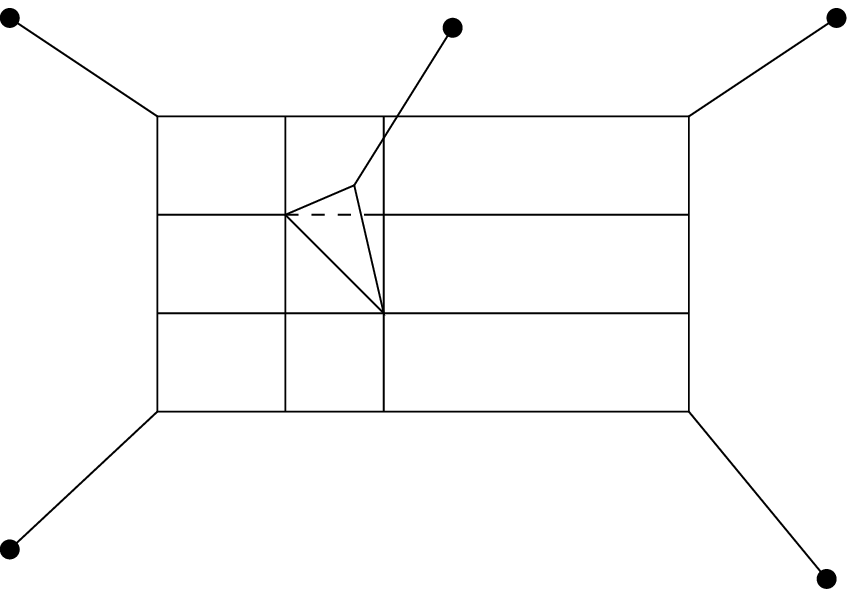} &
    \includegraphics[scale=0.5]{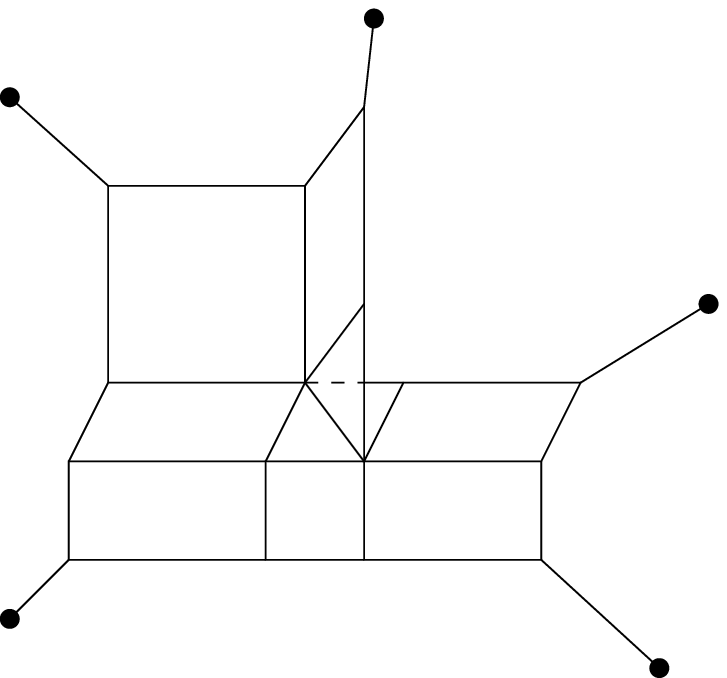} \\
\end{tabular}
\end{center}
\caption{The three canonical types of tight span of 5-point metric space.}
\label{TS_5}
\end{figure}

In general, tight spans are hard to
visualize. Nevertheless, if $|X|\le 5$, Dress \cite{Dr} completely
described $T(X)$ via the interpoint-distances of $X$. For example,
if $|X|=3,$ say $X=\{x,y,z\},$  then $T(X)$ consists of three line
segments joined at a (Steiner) point, with the points of $X$ at the
ends of the arms (see Fig. \ref{TS_3_4} (a)). The lengths of these
segments are $\alpha_{x},\alpha_y,\alpha_z,$ where
$\alpha_x:=(y,z)_x=1/2(d(x,y)+d(x,z)-d(y,z))$ is the Gromov product
of $x$ with the couple $y,z$ ($\alpha_y$ and $\alpha_z$ are defined
in a similar way). Notice that one of the values
$\alpha_x,\alpha_y,\alpha_z$ may be 0, in this case one point is
located between two others. If $|X|=4,$  then the generic form of
$T(X)$ is a rectangle $R(X)$ endowed with the $l_1$-metric, together
with a line segment attached by one end to each corner of this
rectangle (see Fig. \ref{TS_3_4} (b)). The four points of $X$ are the
outer ends of these segments. The lengths of these segments and the
sides of the rectangle can be computed in constant time from the
pairwise distances between the points of $X$;  for exact
calculations see \cite{Dr}. It may happen that $R(X)$ degenerates
into a segment or a point. Finally, there are three canonical types
of tight spans of 5-point metric spaces precisely described in
\cite{Dr} (see also Fig. \ref{TS_5} for an illustration). Each of
them consists of four or five rectangles, five segments, and
eventually one rectangular triangle, alltogether constituting a
2-dimensional cell complex. All sides of the cells can be computed
in constant  time as described in \cite{Dr}. It was also noticed in
\cite{Dr} that if for each quadruplet $X'$ of a finite metric space
$(X,d)$ the rectangle $R(X')$ is degenerated, then $(X,d)$
isometrically embeds into a (weighted) tree and its tight span
$T(X)$ is a tree-network.

From the construction of tight spans of 3- and 4-point metric spaces
immediately follows that any metric space $(X,d)$ with at most 4
points and its tight span $(T(X),\rho)$ can be isometrically
embedded into the $l_1$-plane as shown in Fig. \ref{TS_3_4} (b). This is
no longer true for metric spaces on 5 points:
to embed, some cells of the tight span must be degenerated. If
$|X|=4$ and the rectangle $R(X)$ is non-degenerated, one can easily
show that $R(X)$ isometrically embeds  into the $l_1$-plane only as
an axis-parallel rectangle. Therefore, if additionally the four line
segments of $T(X)$ are also non-degenerated, then up to a rotation
of the plane by $90^{\circ},$ $X$ and $T(X)$  admit exactly two
isometric embeddings into the $l_1$-plane. If one corner of $R(X)$
is a point of $X$  and the embedding of the rectangle $R(X)$ is
fixed, then there exist three types of isometric embeddings of $X$
and $T(X)$ into the rectilinear plane: two segments of $T(X)$ are
embedded as axis-parallel segments and the third one as a segment
whose slope has to be determined. Analogously, if two incident
corners of $R(X)$ are points of $X,$ the two segments of $T(X)$ are
either embedded as axis-parallel segments, or one as a horizontal or
vertical segment and another one as segment whose slope has to be
determined. Note also that from the combinatorial characterization
of finite metric subspaces of the $l_1$-plane presented in
\cite{BaCh_six} immediately follows that a tree-metric $(X,d)$ is
isometrically embeddable into the $l_1$-plane if and only if the
tree-network $T(X)$ has at most four leaves. Finally note that since
$({\mathbb R}^2,d_1)$ is injective, by minimality property of tight
spans, $T(X)$ is an isometric subspace of the $l_1$-plane for any
finite subspace $X$ of ${\mathbb R}^2$.

\section{Algorithm and its correctness}

\subsection{Outline of the algorithm} Let $(X,d)$ be a metric space with $n$ points, called {\it terminals}. Set $X=\{x_1,\ldots,x_n\}$.
Our algorithm first finds in $O(n^2)$ time a
quadruplet $P^{\circ}$ of $X$ whose tight span contains a nondegenerated rectangle $R(P^{\circ})$. If such a quadruplet
does not exists, then $(X,d)$ is a tree-metric and $T(X)$ is a tree-network.  If this
tree-network contains more than four leaves, then $(X,d)$ cannot be isometrically embedded into the $l_1$-plane,
otherwise such an embedding can be easily derived. Given the required
quadruplet $P^{\circ},$ we consider any isometric embedding  of $P^{\circ}$ and of its tight span into the $l_1$-plane as illustrated in Fig. \ref{TP*}
and partition the
remaining points of $X$ into groups depending on their location in the regions of the plane defined by the rectangle $R(P^{\circ})$ and the segments
of $T(P^{\circ})$. The exact location of points of $X$ in these regions is uniquely
determined except the four quadrants defined by $R(P^{\circ})$. At the second stage, we replace the quadruplet $P^{\circ}$ by another quadruplet $P$  by picking one furthest
from $R(P^{\circ})$ point of $X$ in each of these quadrants.  We show that the rectangle $R(P^{\circ})$ is contained in the rectangle $R(P),$ moreover,
for any isometric embedding $\varphi_0$ of $P$ and $T(P)$ into the $l_1$-plane,
the quadrants defined by two opposite corners
are empty (do not contains other terminals of $X$).  Again the location of the points of $X$ in all regions of the plane except the two opposite
quadrants is uniquely determined. To compute the location of the remaining terminals in these two quadrants we adapt the second part of the algorithm of
Edmonds \cite{Ed}: we construct on these terminals a graph as in \cite{Ed}, partition it into connected components, separately determine the location of
the points of each component, and then combine them into
a single chain of components in order to obtain a global isometric embedding $\varphi$ of $(X,d)$ extending $\varphi_0$ or to decide that it does not exist.

Now, we briefly overview the algorithms of Edmonds \cite{Ed} and
Eppstein \cite{Epp}. Edmonds \cite{Ed} starts by picking two
diametral points $p,q$ of $X.$ These two points can be embedded into
the $l_1$-plane in an infinite number of different ways. Each
embedding defines  an axis-parallel rectangle $\Pi$  whose
half-perimeter  is exactly $d(p,q).$ Using the distances of $p$ and
$q$ to the remaining points of $X,$ Edmonds computes a list $\Delta$
of linear size of possible values of the sides of the rectangle
$\Pi.$ For each value $\delta$ from this list, the algorithm of
\cite{Ed} decides in $O(n^2)$ time if there exists an isometric
embedding of $X$ such that one side of the rectangle $\Pi$ has
length $\delta$. For this, it partitions the points of $X$ into
groups, depending on their location in the regions of the plane
determined by $\Pi.$ In order to fix the positions of points in one
of these regions, Edmonds \cite{Ed} defines a graph whose connected
components are also used  in our algorithm. While sweeping through
the list $\Delta,$ the algorithm of \cite{Ed} update this graph and
its connected components in an efficient way. Notice that the second
part of our algorithm is similar to that from \cite{Ed}, but
instead of trying several sizes of the rectangle $\Pi,$ we use the
tight spans to provide us with a single rectangle, ensuring some
rigidity in the embedding of the remaining  points. The algorithm of
Eppstein \cite{Epp} is quite different in spirit from our algorithm
and that of Edmonds \cite{Ed}. Eppstein \cite{Epp} first
incrementally constructs  in $O(n^2)$ time a planar rectangular
complex which is the tight span of the input metric space $(X,d)$ or
decide that the tight span of $X$ is not planar. In the second stage
of the algorithm, he decides in $O(n^2)$ time if this planar
rectangular complex can be isometrically embedded into the
$l_1$-plane or not.

\subsection{Computing the quadruplet $P^{\circ}$}\label{Pcirc} For each $i=1,\ldots,n,$ set $X_i:=\{x_1,\ldots,x_{i}\}.$ We start by computing the tight span of the first four points of $X.$ If this tight span is not degenerated then we return the quadruplet $X_4$ as $P^{\circ}.$ Now suppose that the tight span of the first $i-1$ points of $X$ is a tree-network $A_{i-1}$
with at most four leaves. This means that $A_{i-1}$ contains one or two ramification points (which are not necessarily points of $X$) having degree at most 4,
all remaining terminals of $X_{i-1}$ are either leaves or vertices of degree two of $A_{i-1}.$ We say that two terminals of $X_{i-1}$ are consecutive in $A_{i-1}$
if the segment connecting them in $A_{i-1}$ does not contain other points of $X_{i-1}.$ Note that $A_{i-1}$ contains  at most $n+4$ of consecutive
pairs. For each pair $x_j,x_k$ of consecutive terminals of $X_{i-1}$ we compute the Gromov product $\alpha_{x_i}:=(x_j,x_k)_{x_i}=1/2(d(x_i,x_j)+d(x_i,x_k)-d(x_j,x_k))$ of $x_i$ with $\{ x_j,x_k\}.$ Let $\{ a,b\}$ be the pair of consecutive points of $X_{i-1}$ minimizing the Gromov product $\alpha_{x_i}=(a,b)_{x_i}.$ Let $c$ be the point of the segment $[a,b]$ of $A_{i-1}$ located at distance $\alpha_a:=(b,x_i)_a$ from $a$ and at distance $\alpha_b:=(a,x_i)_b$ from $b$ ($c$ may coincide with one of the points $a$ or $b$).

\begin{figure}
\begin{center}
    \includegraphics[scale=0.5]{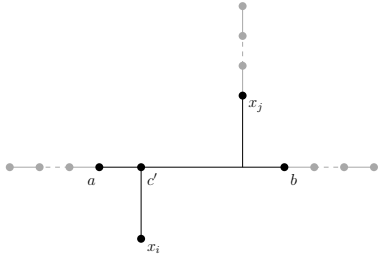}
\end{center}
\caption{The tree-network $A_i.$}
\label{tree_A_i}
\end{figure}

Denote by $A_i$ the tree-network obtained from $A_{i-1}$ by adding the segment $[x_i,c]$ of length $\alpha_{x_i}.$ By running  Breadth-First-Search on $A_i$  rooted at $x_i,$  we check if $d_{A_i}(x_i,x_j)=d(x_i,x_j)$ for any terminal $x_j$ of $X_{i}.$ If this holds for all $x_j\in X_{i},$ then the tight span of $X_i$ is the tree-network $A_i.$ If $A_i$ contains more than 4 leaves, then we return the answer ``not''and the algorithm halts. Otherwise, if $i=n,$ then we return the answer  ``yes'' and an isometric embedding of $X$ and its tight
span $A_n$ in the $l_1$-plane, else, if $i<n,$  we consider the next point $x_{i+1}.$ Finally, if $x_j$ is the first point of $X_i$ such that $d_{A_i}(x_i,x_j)\ne d(x_i,x_j),$
then we assert that {\it the tight span of the quadruplet $\{ a,b,x_i,x_j\}$ is non-degenerated and we return it as $P^{\circ}.$} Suppose by way of contradiction that $T(P^{\circ})$ is a tree.
Since $A_{i-1}$ realizes $X_{i-1}$ and $T(P^{\circ})$ realizes
$P^{\circ}$, the subtree of $A_{i-1}$ spanned by the terminals
$a,b,$ and $x_j$ is isometric to the subtree of $T(P^{\circ})$
spanned by the same terminals. On the other hand, $T(P^{\circ})$ contains a point $c'$ located at
distance $\alpha_{x_i},\alpha_a,$ and $\alpha_b$ from
$x_i,a,$ and $b,$ respectively.
This means that $T(P^{\circ})$ is isometric to the subtree of $A_i$
spanned by the vertices $x_i,a,b,$ and $x_j,$ (see Fig.~\ref{tree_A_i}) contrary to the
assumption that $d_{A_i}(x_i,x_j)\ne d(x_i,x_j).$
Hence, this inequality implies indeed that $T(P^{\circ})$ is not a tree. Finally note that dealing with a current
point $x_i$ takes time linear in $i,$ thus the whole
algorithm for computing the quadruplet $P^{\circ}$ runs in $O(n^2)$ time.

\subsection{Classification of the points of $X$ with respect to the rectangle of $T(P^{\circ})$}\label{Pcircb} Let $P^{\circ}=\{ p^{\circ}_1,p^{\circ}_2,p^{\circ}_3,p^{\circ}_4\}$
be the quadruplet whose tight span $T(P^{\circ})$ is non-degenerated.  Let
$R^{\circ}$ be one of the two possible isometric embeddings of the rectangle $R(P^{\circ})$ of $T(P^{\circ})$ and consider a complete or a partial isometric embedding of $T(P^{\circ})$ such that $R(P^{\circ})$ is embedded as $R^{\circ}.$
Denote by
$Q^{\circ}_1,Q^{\circ}_2,Q^{\circ}_3,Q^{\circ}_4$ the four (closed) quadrants defined by the four consecutive corners $q^{\circ}_1,q^{\circ}_2,q^{\circ}_3,q^{\circ}_4$ of $R^{\circ}$ labeled in such a way that
the point $p^{\circ}_i$ must be located in the quadrant $Q^{\circ}_i, i=1,\ldots,4.$ Let also $S^{\circ}_1,S^{\circ}_2,S^{\circ}_3,$ and $S^{\circ}_4$ be the remaining half-infinite strips.
Since we know how to construct in constant time the tight span of  a 5-point
metric space, we can compute the distances from all terminals $p$ of $X$ to the corners of the rectangle $R(P^{\circ})$ (and hence to the corners of $R^{\circ}$) in total $O(n)$ time. With some abuse of notation,
we will denote the $l_1$-distance from $p$ to the corner $q^{\circ}_i$ of $R^{\circ}$ by $d(p,q^{\circ}_i).$
Since $R^{\circ}$ is gated,
from the distances of $p$ to the corners  of $R^{\circ}$ we can compute the gate of  $p$ in $R^{\circ}.$ Consequently, for each point $p\in X\setminus P^{\circ}$ we can decide in which of
the nine regions of the plane will belong its location $\varphi(p)$ under any isometric embedding $\varphi$ of $(X,d)$ subject to the assumption that $R(P^{\circ})$ is embedded as $R^{\circ}$.
If $\varphi(p)$ belongs to one of the four half-strips or to $R^{\circ}$, then we can also easily find
the exact location itself: this can be done by using either the gate of $p$ in $R^{\circ}$ or the fact that inside these five regions the intersection of the four $l_1$-spheres centered at the corners of $R^{\circ}$ and having the distances from respective corners to $p$ as radii is a single point. So, it remains to decide the locations
of points assigned to the four quadrants $Q^{\circ}_1,Q^{\circ}_2,Q^{\circ}_3,$ and $Q^{\circ}_4.$  For any point $p\in X$
which must be located in the quadrant $Q^{\circ}_i,$ the set of possible locations of $p$ is either empty (and no isometric embedding exists) or a segment $s_p$ of $Q^{\circ}_i$ consisting of all points $z\in Q^{\circ}_i$ such that $\|z-q^{\circ}_i\|_1=d(p,q^{\circ}_i).$

Notice that for any quadruplet $P'=\{ p'_1,p'_2,p'_3,p'_4\}$ of terminals such that $p'_i$ is assigned to the quadrant $Q^{\circ}_i,$ $i\in \{1,2,3,4\},$ {\it the rectangle $R^{\circ}$ belongs to the tight span $T(P')$ of $P'.$} Indeed, for any point $p'_i, i\in \{ 1,2,3,4\}$ and any point $r$ of $R^{\circ},$ we have $\|p'_i-r\|_1+\|r-p'_j\|_1=\|p'_i-p'_j\|_1,$ where $j$ is selected in such a way that $q^{\circ}_i$ and $q^{\circ}_j$ are opposite corners of $R^{\circ}.$ From injectivity of the $l_1$-plane and the characterization of tight spans we conclude that all points of $R^{\circ}$ belong
to $T(P'),$ establishing in particular that this tight span is also non-degenerated.

\subsection{The quadruplet $P$ and its properties}\label{P} Let $P=\{ p_1,p_2,p_3,p_4\}$
be the quadruplet of $X,$ where $p_i$ is a point of $X$ which must
be located in the quadrant $Q^{\circ}_i$ and is maximally distant
from the corner $q^{\circ}_i$ of $R^{\circ}$. As we established
above, the tight span of $P$ is non-degenerated,  moreover the
rectangle $R(P)$ contains the rectangle $R(P^{\circ}).$ As we also
noticed, there exists a constant number of ways in which we can
isometrically embed  $T(P)$ into the $l_1$-plane. Further we proceed
in the following way: we pick an arbitrary isometric embedding
$\varphi_0$ of $T(P)$ and try to extend it to an isometric embedding
$\varphi$ of the whole metric space $(X,d)$ in the $l_1$-plane. If
this is possible for some embedding of $T(P)$, then the algorithm
returns the answer ``yes'' and an isometric embedding of $X,$
otherwise the algorithm returns the answer ``not''. Let $R$ denote
the image of $R(P)$ under $\varphi_0.$

\begin{figure}
\begin{center}
\includegraphics[scale=0.5]{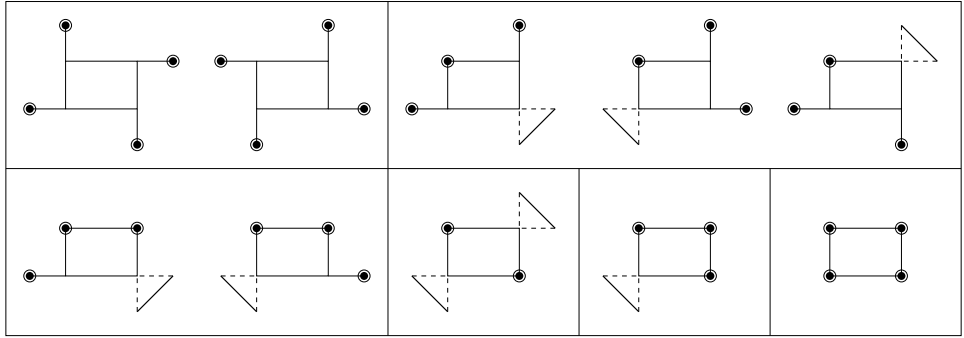}
\end{center}
\caption{Possible isometric embeddings of $T(P).$}
\label{TP*}
\end{figure}

We call a terminal $p_i$ of $P$ {\it fixed} by the embedding $\varphi_0$ if either $\varphi_0(p_i)$
is a corner of the rectangle $R$ or the segment of $T(P)$ incident to $p_i$ is embedded by
$\varphi_0$ as a horizontal or a vertical segment; else we call $p_i$ {\it free}.
The embedding of a free terminal $p_i$ is not exactly determined but is restricted
to a segment $s_{p_i}$ consisting  of the points of the quadrant defined by $q_i$
and having the same $l_1$-distance to $q_i.$ We call the terminals
$p_i,p_{i+1(\mbox{\footnotesize mod} 4)}$  {\it incident} and
the terminals $p_i,p_{i+2(\mbox{\footnotesize mod} 4)}$ {\it opposite}.
From the isometric embedding of $T(P)$ we conclude that at most one of
two incident terminals can be free. Moreover, if a terminal $p_i$ of $P$ is
fixed but is not a corner of $R,$ then at least one of the two terminals
incident to $p_i$ is also fixed. If all four tips of $T(P)$ are non-degenerated,
then all four terminals of $P$ are fixed. If only three tips of $T(P)$ are
non-degenerated then at most one terminal of $P$ is free, all remaining
terminals are fixed. If only two tips of $T(P)$ are non-degenerated,
then either they correspond to incident terminals, one of which is
fixed and another one is free or to two opposite terminals which are both free.
Finally, if only one tip of the tight span is non-degenerated, then it corresponds
to a free terminal, all other terminals of $P$ are corners of $R$ and therefore
are fixed (see Fig. \ref{TP*} for the occurring  possibilities).

\begin{figure}
\begin{center}
\includegraphics[scale=0.6]{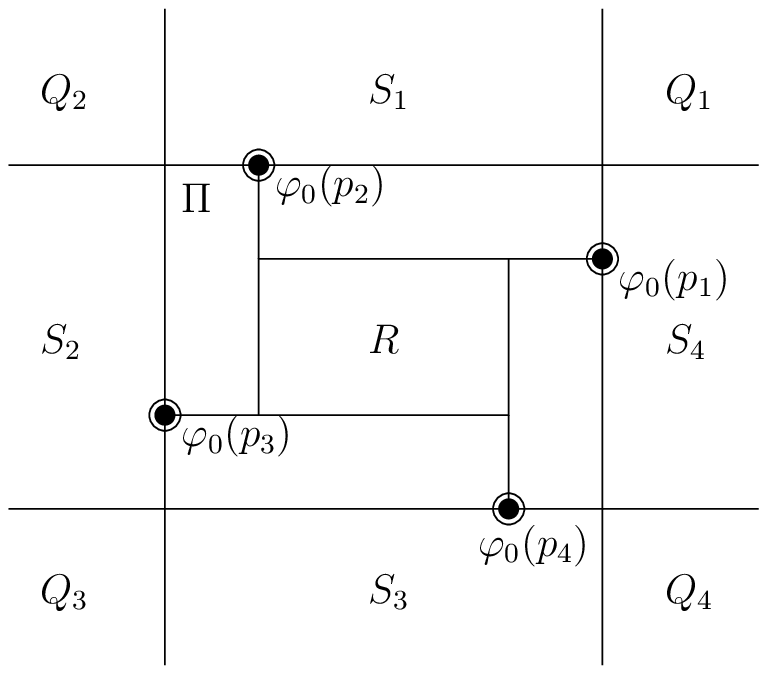}
\hspace*{2cm}
\includegraphics[scale=0.6]{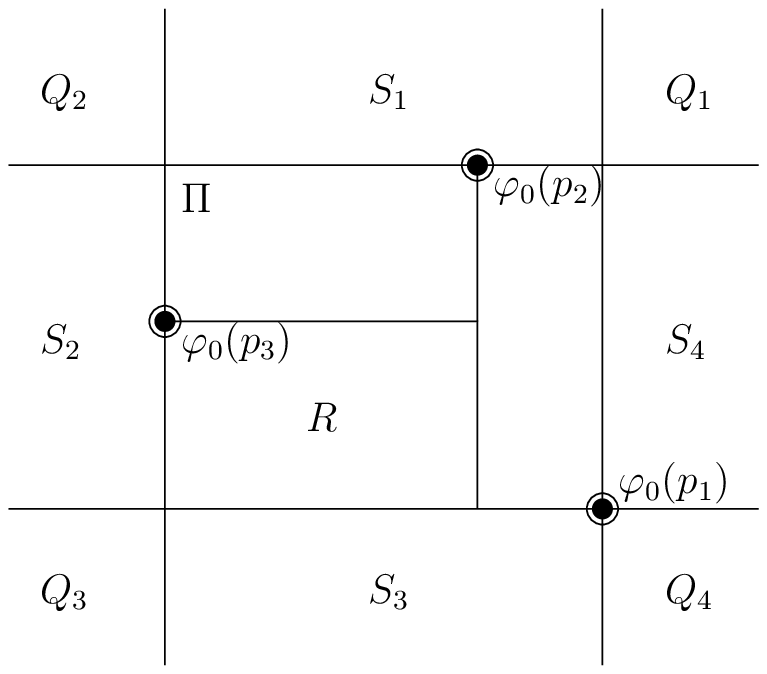}
\end{center}
\caption{The partition of the plane into half-strips and quadrants.}
\label{regions}
\end{figure}

 Denote by $\Pi$ the smallest axis-parallel rectangle containing $R$ and the fixed terminals
 of $P;$ Fig. \ref{regions} illustrates $\Pi$ for two cases from Fig. \ref{TP*}
 (if a terminal is free, then the respective corner of $R$ is also a corner of
 $\Pi$). Let $q_1,q_2,q_3,q_4$ be the corners of $\Pi$ labeled in such a way that $q_i$ is the corner of
$R$ corresponding to the point $p_i$ and to the corner $q^{\circ}_i$
of $R^{\circ}.$ Denote by $Q_1,\ldots,Q_4$ the quadrants of
${\mathbb R}^2$ defined by
 the corners of $\Pi$ and by $S_1,\ldots,S_4$ the remaining half-infinite strips.
 Again, as in the case of the quadruplet $P^{\circ}$, by building the tight spans
 of $P\cup \{ p\}$ for all terminals $p\in X\setminus P,$ we can compute in total
 linear time the distances from all such points $p$ to the corners of $R$ (and to
 the corners of $\Pi$). From these four distances and the distances of $p$ to the
 terminals of the quadruplet $P$ we can determine in which of the nine regions
 $Q_1,Q_2,Q_3,Q_4,S_1,S_2,S_3,S_4,\Pi$ of the plane must be located $p.$
 Moreover, if $p$ is assigned to the rectangle $\Pi$ or to one of the four
 half-strips $S_1,S_2,S_3,S_4,$ then we can conclude that, in the region 
 in which $p$ assigned, the intersection
 of the four spheres centered at the terminals of $P$ and having the distances
 from respective points to $p$ as radii is either empty or a single point. The
 sphere centered at a free terminal $p_i$ is needed only to decide the location of
 $p$ in the quadrant $Q$ of the plane having the same apex $a$ as the quadrant $Q_i$
 and which is opposite to  $Q_i$ ($a$ is a corner of $\Pi$). But in this case,
 instead of considering the sphere of radius $d(p,p_i)$ centered at $\varphi_0(p_i)$
 we consider the sphere of radius $d(p,p_i)-\|\varphi_0(p_i)-a\|_1$ and centered
 at $a:$ indeed, both these spheres have the same intersection with $Q$.

We are now ready to prove the following property of the quadruplet $P:$ {\it among the four quadrants
$Q_1,Q_2,Q_3,$ and $Q_4$ defined by $P,$ two opposite quadrants are empty,} i.e., they do not contain terminals
of $X\setminus P$. First note that by inspecting the different cases listed in Fig.~\ref{TP*} one can check that the
two neighbors $p_{i-1(\mbox{\footnotesize mod} 4)}$ and $p_{i+1(\mbox{\footnotesize mod} 4)}$ of a free point $p_i\in P$ are both
fixed; let say $p_1$ and $p_3$ are fixed. Now, suppose by way of contradiction that a terminal $q\in X\setminus P$ must be located
in the quadrant $Q_1.$ This means that its gate in the rectangle $\Pi$ is the corner of $\Pi$ corresponding to $p_1$. Since in any embedding $\varphi$ of $X$
that extends the chosen embedding of $T(P^{\circ})$ the terminal $p_1$ is located in $Q^{\circ}_1,$ we deduce that  $Q_1(\varphi(p_1))\subseteq  Q^{\circ}_1.$
On the other hand, the inclusion $Q_1\subseteq Q_1(\varphi(p_1))$ follows directly from the definition of $Q_1$ and the fact
that $p_1$ is fixed. Now, from the inclusions $Q_1\subseteq Q_1(\varphi(p_1))\subseteq Q_1^{\circ}$, we obtain that $q\in Q_1^{\circ}$ and,
since $q$ is closer to $p_1$ than to $q^{\circ}_1,$ we get a contradiction with the choice of $p_1,$
establishing that indeed $Q_1$ does not contain any point of $X\setminus P.$ The same argument
shows that $Q_3$ is empty as well. Note that actually we proved that any quadrant $Q_i$ corresponding to a
fixed terminal $p_i$ of $P$ is empty.

\begin{figure}
\begin{center}
\includegraphics[scale=0.6]{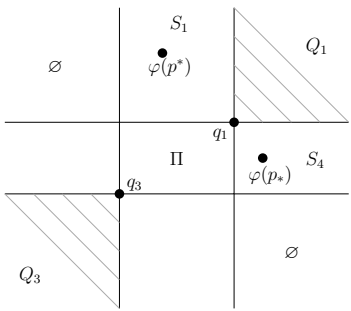}
\end{center}
\caption{On possible locations of terminals in $Q_1$ and $Q_3.$}
\label{zone}
\end{figure}


\subsection{Locating in the non-empty quadrants $Q_1$ and $Q_3$}\label{Q}

As we have showed in previous subsection, any isometric embedding $\varphi$ of $(X,d)$ extending the embedding $\varphi_0$ of $T(P)$ locates each terminal $p$ of $X\setminus P$
in one and the same of the nine regions defined by $\Pi.$ Moreover, if $p$ must be located in the rectangle $\Pi$ or in one of the four half-strips $S_1,\ldots,S_4,$
then this location $\varphi(p)$ is uniquely determined from the distances to the terminals of $P$ and to the corners of $\Pi.$ We also established that
only one or two opposite quadrants defined by $\Pi$, say $Q_1$ and $Q_3,$ can host terminals of $X\setminus P;$ see Fig. \ref{zone}. We will show now how
to find the exact location of the set $X_1$ of terminals assigned to $Q_1$ (the set $X_3$ of terminals which must be located in $Q_3$ is treated analogously).

Note that independently of how the extension $\varphi$ of $\varphi_0$ is chosen, for each terminal $u\in X_1,$ the $l_1$-distance $\|\varphi(u)-q_1\|_1$ from the location of $u$ to the corner $q_1$ of $\Pi$ is one and the same, which we denote by $\Delta_u.$ The value of  $\Delta_u$  can be easily computed because $q_1$ lies between $\varphi(u)$ and $\varphi(p_i)$ for any $p_i\in P$: for example, we can set $\Delta_u:=d(u,p_1)-\| \varphi_0(p_1)-q_1\|_1.$   Then the set of all possible locations $\varphi(u)$ of $u\in X_1$ is the {\it level segment} $s_u$ which is the intersection of $Q_1$  with the sphere $S(q_1,\Delta_u)$ of radius  $\Delta_u$ centered at $q_1.$

To compute the locations of the terminals of $X_1$ in the quadrant $Q_1$, we adapt to the $l_1$-plane
the definition of a graph (which we denote by $G_1=(X_1,E_1)$) defined by Edmonds \cite{Ed} in the $l_{\infty}$-plane.
Two terminals $u,v \in X_1$ are adjacent in $G_1$ if and only if  $d(u,v)>|\Delta_u-\Delta_v|.$ Equivalently
$u,v\in X_1$ with $\Delta_u\le \Delta_v$ are adjacent in $G_1$ iff $u$ cannot be located between $q_1$ and
$v:$ $\varphi(u)\notin I_1(q_1,\varphi(v)).$  Denote by $C_1,C_2,\ldots,C_k$ the connected components of the graph $G_1.$
They have the following useful properties established in Lemmata 3-5 of \cite{Ed}:

\medskip\noindent
(1) Each component $C_i$ is {\it rigid,} i.e., once the location
of any point $u$ of  $C_i$ has been fixed, the locations of the remaining points of $C_i$
are also fixed (up to symmetry with respect to the line parallel to the bisector of  $Q_1$ and
passing via $u$);

\medskip\noindent
(2) The components $C_1, C_2, \ldots, C_k$ of the graph $G_1$ can be numbered so that the
points of each $C_i$ appear consecutively in the list of points  $u\in X_1$
sorted in increasing order of their distances $\Delta_u$ to $q_1;$

\medskip\noindent
(3) For a component $C_i$ of $G_1,$ let ${B}_i$ be the smallest
axis-parallel rectangle containing $\{\varphi_i(u): u\in C_i\}$ for an isometric embedding $\varphi_i$
of $(C_i,d)$ in the $l_1$-plane. Let $b_i$ be the upper right corner
of ${B}_i.$ Then the embedding of $C_1, C_2,\ldots, C_k$ preserves
the distances between all pairs of points lying in different
components if and only if for every pair of consecutive components
$C_i$ and $C_{i+1},$ the rectangle ${B}_{i+1}$ lies entirely in the
quadrant $Q_1(b_i).$

\begin{figure}
    \begin{minipage}[b]{.48\linewidth}
        \centering
        \includegraphics[scale=0.5]{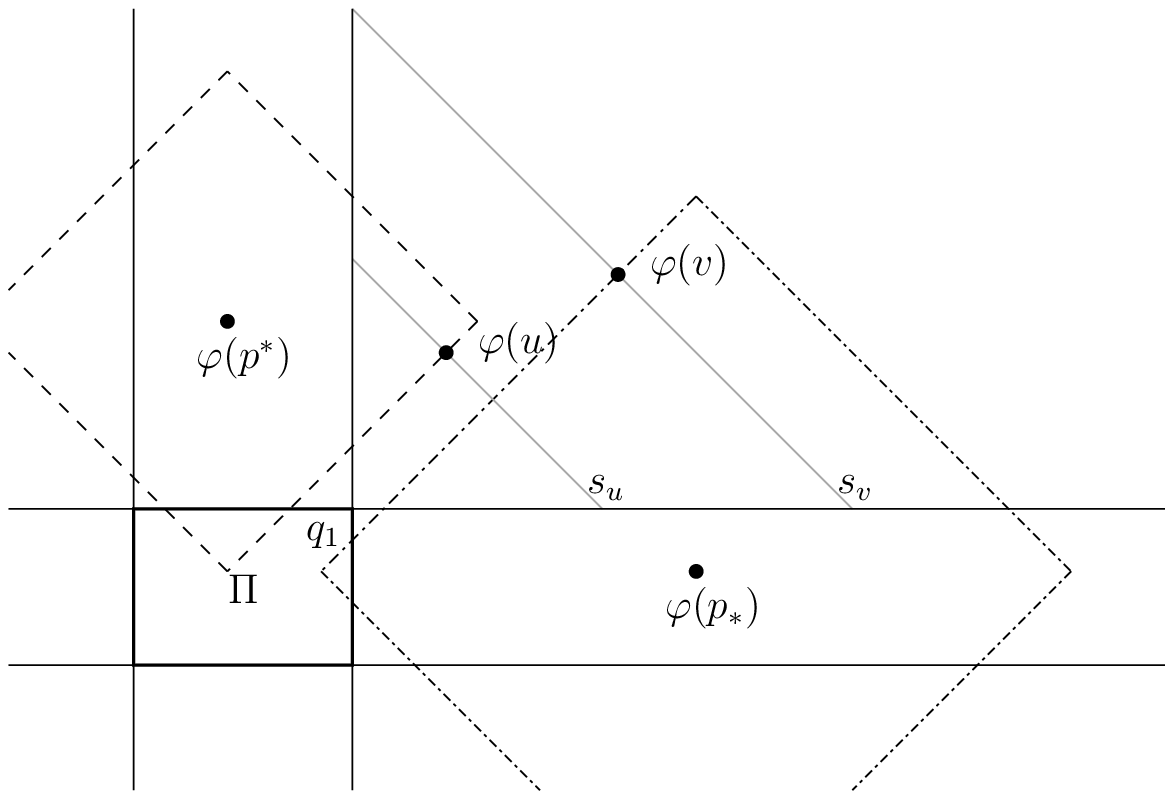}
        \caption{$\varphi(u)$ and $\varphi(v)$ are fixed by $\varphi(p^*)$ and $\varphi(p_*)$}
        \label{points_fixes}
    \end{minipage} \hfill
    \begin{minipage}[b]{.48\linewidth}
        \centering
        \includegraphics[scale=0.5]{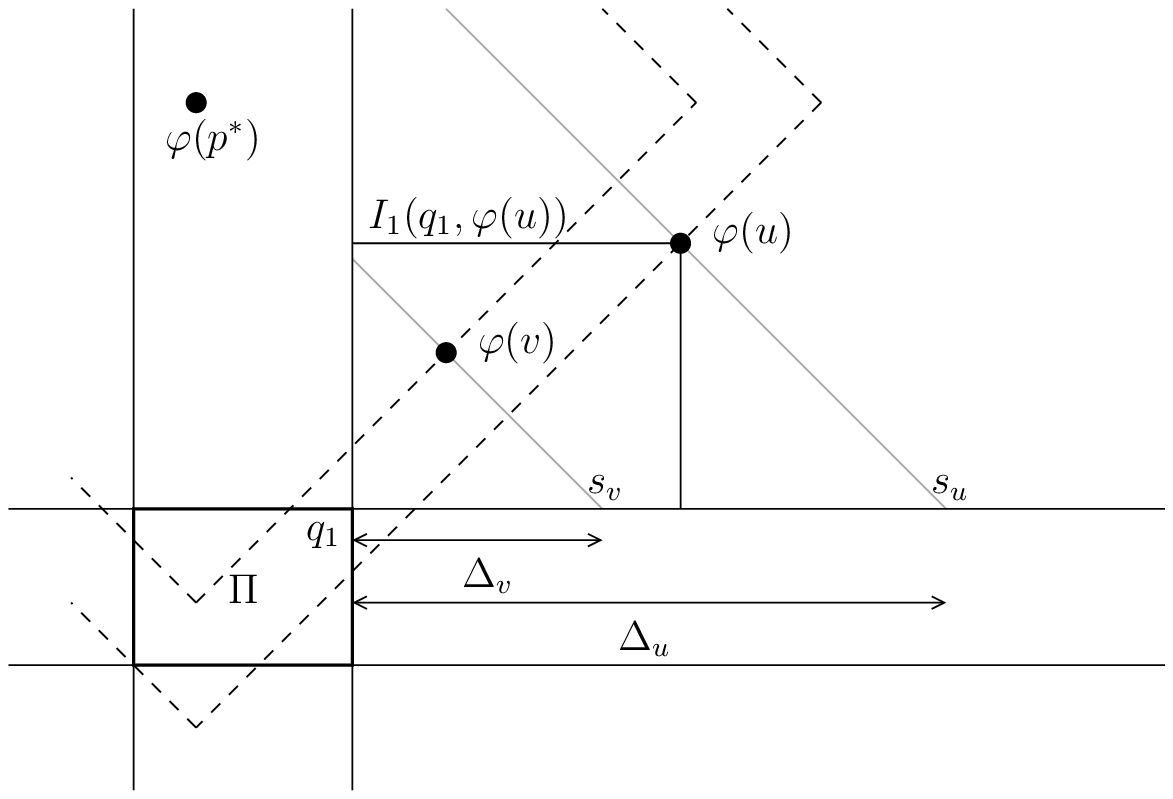}
        \caption{$\varphi(v)$ is fixed by $\varphi(p^*)$}
        \label{fixe_v}
    \end{minipage}
\end{figure}

\medskip
The location in the quadrant $Q_1$ of some terminals of $X_1$ (and
therefore of the connected components containing them) can be fixed
by terminals already located in the two half-strips incident to
$Q_1.$ We say that a terminal $u\in X_1$ is {\it fixed by a
terminal} $p$ already located in $S_1\cup S_4$ if the intersection
of the segment $s_u$ with the sphere $S(\varphi(p),d(p,u))$ is a
single point. Note that if $u\in X_1$ is fixed by a terminal located
in $S_1,$ then $u$ is also fixed by the upmost terminal $p^*$
located in this half-strip. Analogously, if $u\in X_1$ is fixed by a
terminal of  $S_4,$ then $u$ is also fixed by the rightmost terminal
$p_*$ located in $S_4.$ Therefore by considering the intersections
of the segments $s_u, u\in X_1,$ with the spheres
$S(\varphi(p^*),d(p^*,u))$ and $S(\varphi(p_*),d(p_*,u))$ we can
decide in linear time which terminals of $X_1$ are fixed by $p^*$
and $p_*$ and find their location in $Q_1$ (for an illustration, see
Fig. \ref{points_fixes}). According to property (1), if a terminal
of a connected component of $G_1$ is fixed, then the location of the
whole component is also fixed (up to symmetry).  Let $C_j$ be the
connected component of $G_1$ containing the furthest from $q_1$
terminal $u\in X_1$ fixed by $p^*$ or $p_*,$ say by $p^*$ (therefore
the location of $C_j$ is fixed). We assert that {\it  all terminals
of $C_1, C_2, ..., C_{j-1}$ are also fixed by $p^*$.} Indeed, pick
such a terminal $v.$ From property (2) we conclude that $\Delta_v\le
\Delta_u$ and from the definition of $G_1$ we deduce that $v$ must
be located in the axis-parallel rectangle $I_1(q_1,\varphi(u)),$ and
therefore below $u$. Since $u$ is fixed by $p^*,$ $u$ must be
located below $p^*$, whence $v$ also must be located below $p^*.$ We
can easily see that the intersection of $s_v$ with the sphere
$S(\varphi(p^*),d(p^*,v))$ is a single point, i.e. $v$ is also fixed
by $p^*$  (see Fig. \ref{fixe_v}).

\begin{figure}
    \begin{center}
    \includegraphics[scale=0.6]{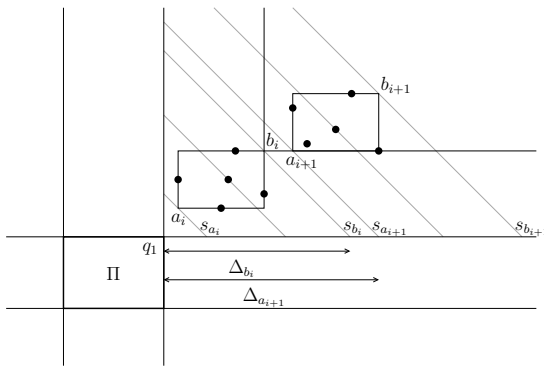}
    \end{center}
    \caption{On the assemblage of blocks $B_{j+1},\ldots, B_k.$}
    \label{blocs}
\end{figure}

It remains to locate in $Q_1$ the terminals of the 
components $C_{j+1}, C_{j+2},\ldots,C_k.$ We compute separately an
isometric embedding of each component $C_i$ for $i=j+1,\ldots,k.$
For this, we fix arbitrarily the location of the first two
points $u,v$ of $C_i$ in the segments $s_u$ and $s_v$ so that to
preserve the distance $d(u,v)$ (the terminals of $C_i$ are ordered
by their distances to $q_1$). By property (1) of \cite{Ed}, the
location of the remaining points of $C_i$ is uniquely determined and
each  point $w$ of $C_i$ will be located in its level segment $s_w.$
Let $\varphi_i$ be the resulting embedding of $C_i$.  Denote by
${B}_i$ the smallest axis-parallel rectangle (alias {\it box})
containing the image $\varphi_i(C_i)$ of $C_i.$ Let $a_i$ and $b_i$
denote the lower left and the upper right corners of  ${B}_i.$ Note
that $a_i$ belongs to the $l_1$-interval between $q_1$ and the image
$\varphi_i(u)$ of any terminal $u$ of $C_i,$ while the
$l_1$-interval between $q_1$ and $b_i$ will contain the images of
all terminals of $C_i.$ Therefore if we set
$\Delta_{a_i}:=\Delta_u-\|a_i-\varphi_i(u)\|_1$ and
$\Delta_{b_i}:=\Delta_u+\| \varphi(u)-b_i\|_1,$ where $u$ is any
terminal of $C_i,$ then in all isometric embeddings of $(C_i,d)$ in
which all terminals $u\in C_i$ are located on $s_u,$ the points
$a_i$ and $b_i$ must be located on the level segments $s_{a_i}$ and
$s_{b_i},$  defined as the intersections of the quadrant $Q_1$ with
the spheres $S(q_1, \Delta_{a_i})$ and $S(q_1,\Delta_{b_i}).$

By properties (2) and (3) of \cite{Ed}, in order to define a single
isometric embedding of the components $C_{j+1},\ldots,C_k$ we now
need to assemble the boxes ${B}_{j+1},\ldots, {B}_k$ (by moving
their terminals along the level segments) in such a way that {\it
for two consecutive components $C_i$ and $C_{i+1},$ the box
${B}_{i+1}$ lies entirely in the quadrant $Q_1(b_i).$}  We assert
that this is possible if and only if {\it for each pair of
consecutive boxes ${B}_i,{B}_{i+1},$ $i=j,j+1, ...,k-1,$ the
inequality $\Delta_{b_i}\le \Delta_{a_{i+1}}$ holds.}  Indeed, if
$\Delta_{b_i}\le \Delta_{a_{i+1}},$ then translating  ${B}_{i+1}$
along the segment $s_{a_{i+1}},$ we can locate its corner $a_{i+1}$
in the quadrant $Q_1(b_i)$ and thus satisfy the embedding
requirement.  Conversely, if $\Delta_{b_i}>\Delta_{a_{i+1}}$ holds,
then  $a_{i+1}$ cannot belong to the quadrant $Q_1(b_i)$
independently of the positions of $a_{i+1}$ and $b_i$ on their level
segments. This local condition depends only of the values of
$\Delta_{a_i}, \Delta_{b_i}$ and is independent of the actual
location of the boxes $B_{i},i=1,\ldots,k.$  As a result, the
algorithm that embeds the boxes ${B}_{j+1}, ..., {B}_k$ is very
simple. For each $i=j,...,k-1,$ we compute the box ${B}_{i+1}$ and
the values of $\Delta_{a_{i+1}}$ and $\Delta_{b_{i+1}}.$ If
$\Delta_{a_{i+1}}<\Delta_{b_{i}}$ for some $i,$ then return the
answer ``there is no isometric embedding of $(X,d)$ extending the
embedding $\varphi_0$ of $T(P)$". Otherwise, having already located
the box $B_i,$ by what has been shown above, the intersection of the
quadrant $Q_1(b_i)$ with the level segment $s_{a_{i+1}}$ is
non-empty. Therefore we can translate ${B}_{j+1}$ in such a way that
its lower left corner $a_{i+1}$ becomes a point of this
intersection.

In this way, we obtain an embedding of $C_{j+1},\ldots,C_k$ and
$B_{j+1},\ldots,B_{k}$ satisfying the conditions (1)-(3), thus an
isometric embedding of the metric space $(\bigcup_{i=j+1}^kC_i,d)$
in $Q_1.$ Analogously, by constructing the graph $G_3=(X_3,E_3)$ and
its components, either we obtain a negative answer or we return an
isometric embedding  of the metric space defined by the non-fixed
components of $G_3$ in the quadrant $Q_3.$ Denote by $\varphi$ the
embedding of $X$ which coincides with $\varphi_0$ on $P,$ with these
two embeddings on the non-fixed components of $G_1$ and $G_3,$ and
with the already computed fixed locations of the terminals assigned
to $\Pi,$ to the half-strips $S_1,S_2,S_3,S_4,$ and to the fixed
connected components of the graphs $G_1$ and $G_3.$ In $O(n^2)$ we
test if $\varphi$ is an isometric embedding of $(X,d)$ into the
$l_1$-plane. If the answer is negative, then we return ``there is no
isometric embedding of $(X,d)$ extending the embedding $\varphi_0$
of $T(P)$", otherwise we return $\varphi$ as an isometric embedding.
The algorithm returns the global answer ``not'' if for all possible
embeddings $\varphi_0$  of $T(P)$ it returns the negative answer.
From what we established follows that in this case $(X,d)$ is not
isometrically embeddable into the $l_1$-plane.

\subsection{Algorithm and its complexity}

We conclude the paper with a description of the main steps of the algorithm and their complexity.

\medskip
{\footnotesize
\noindent{{\sf Algorithm} \bf Embedding into the $l_1$-plane}\\
{\bf Input:} A metric space $(X,d)$ on $n$ points\\
{\bf Output:} An isometric embedding $\varphi$ of $(X,d)$ into $({\mathbb R}^2,d_1)$
or the answer ``not'' if it does not exist\\
{\bf Step 1.}  Find a quadruplet $P^{\circ}$ of $X$ whose tight span contains
a rectangle. If $P^{\circ}$ does not exist, then $T(X)$ is a tree. If $T(X)$
 has more than 4 leaves, then return ``not", else return an embedding of $T(X)$ and $(X,d)$.\\
{\bf Step 2.}  Pick any embedding of $T(P^{\circ})$ and for each terminal of $X\setminus P^{\circ}$
determine in which of the nine regions of the plane it must be located.
  Using this partition of $X\setminus P^{\circ},$ define the quadruplet $P$.\\
{\bf Step 3.}  Embed $P$ and its tight span $T(P)$ into the $l_1$-plane in all
possible different ways.  Try to extend each of these embeddings
  to an isometric embedding of $(X,d)$ following the rules (a)-(g).
  If all of these attempts return the answer ``not'', then return the answer ``not", else return one of the obtained  embeddings.
\begin{itemize}
\item[(a)]  Given an embedding $\varphi_0$ of $T(P),$ for each terminal $u$ of $X\setminus P$ determine in which of the nine regions defined by the rectangle $\Pi$ will be located $u$ in any isometric embedding extending  $\varphi_0;$
\item[(b)]  Locate the terminals assigned to the rectangle $\Pi$ and the four half-strips  $S_1,S_2,S_3,S_4;$
\item[(c)]  Define the sets of terminals $X_1$ and $X_3$ assigned to the quadrants $Q_1$ and $Q_3,$ construct the graphs $G_1=(X_1,E_1)$ and $G_3=(X_3,E_3)$ and their connected components;
\item[(d)]  Find the terminals of $X_1$ fixed by $p^*,p_*$ and their location in $Q_1$. Do a similar thing for $X_3;$
\item[(e)]  Find an isometric embedding of each component $C_i$ of $G_1$ not containing already fixed terminals so that its terminals are located on their level segments. Do a similar thing for $G_3$;
\item[(f)]  Test if the free components $C_{j+1},\ldots, C_k$ of $G_1$ satisfy the condition  $\Delta_{b_i}\le \Delta_{a_{i+1}}$ for $i=j+1,\ldots,k-1.$ If not, then return the answer ``not", else locate consecutively the boxes $B_{j+1},\ldots,B_k$ in such a way that  $a_{i+1}$ is located in $Q_1(b_i)\cap s_{a_{i+1}}$ and fix in this way the position of all terminals of $X_1.$ Do a similar thing for the free components of $G_3;$
\item[(g)]  Verify if the resulting embedding of $X$ extending $\varphi_0$ is an isometric embedding of $(X,d)$. If ``yes'', then return it as a resulting isometric embedding, otherwise return the answer ``there is no isometric embedding of $(X,d)$ extending the embedding $\varphi_0$''.
\end{itemize}
}

\medskip
In Subsection \ref{Pcirc} we established that the quadruplet
$P^{\circ},$ if it exists, can be computed in $O(n^2)$ time. If
$P^{\circ}$ does not exists, then the tree-network $A_n$
(constructed within the same time bounds) is the tight span of
$(X,d).$ Embedding $A_n$ (if it has at most 4 leaves) in the
$l_1$-plane can be easily done in linear time. As shown in
Subsection \ref{Pcircb}, Step 2 can be implemented in linear time.
There exists a constant number of ways in which the quadruplet $P$
and its tight span can be isometrically embedded in the $l_1$-plane.
Therefore, to show that
Step 3 has complexity $O(n^2)$, it suffices to estimate the total
complexity of the steps (a)-(g) for a fixed embedding $\varphi_0$ of
$T(P).$ Step (a) is similar to Step 2, thus its complexity is
linear. The exact location of each terminal in the half-strips or in
$\Pi$ is determined as the intersection of two spheres, therefore
step (b) is also linear. Defining the graph $G_1$ and computing its
connected components can be done in $O(|X_1|^2)$ time. Thus step (c)
has complexity $O(n^2).$ Steps (d) and (e) can be implemented in an
analogous way as (b), thus their complexity is $O(n).$ Testing the
condition in step (f) and assembling the free components into a
single chain is linear as well. Finally, step (g) requires $O(n^2)$
time. Therefore, the total complexity of the algorithm is $O(n^2).$
Summarizing, here is the main result of this note:

\begin{theorem} For a metric space $(X,d)$ on $n$ points, it is possible to decide in optimal $O(n^2)$ time if $(X,d)$ is isometrically embeddable into the $l_1$-plane and to find such an embedding if it exists.

\end{theorem}

\end{document}